\documentclass[twocolumn,preprintnumbers,superscriptaddress,nofootinbib,aps,prd,floatfix]{revtex4-2}

\usepackage{enumerate}
\usepackage{amsmath,amssymb}
\usepackage{graphicx}
\usepackage{slashed}
\usepackage{xspace,slashed}
\usepackage{hyperref}
\hypersetup{colorlinks=true, citecolor=blue, urlcolor=blue, linkcolor=blue}
\usepackage[normalem]{ulem}
\usepackage{subfigure,orcidlink}
\usepackage{multirow,array}
\usepackage{setspace}
\usepackage{enumitem}

\usepackage{siunitx}
\usepackage{changepage}
\usepackage{hyperref}
\hypersetup{colorlinks=true, citecolor=blue, urlcolor=blue, linkcolor=blue}

\def\cosw{c_{\rm w}}

\def\mw{m_{\rm W}}
\def\mz{m_{\rm Z}}
\def\mh{m_{\rm H}}

\newcommand{\nn}{\nonumber}
\newcommand{\be}{\begin{equation}}
\newcommand{\ee}{\end{equation}}
\newcommand{\bear}{\begin{eqnarray}}
\newcommand{\eear}{\end{eqnarray}}
\newcommand{\mL}{\mathcal{L}}
\newcommand{\mO}{\mathcal{O}}

\def\1loop{one-loop}

\newcommand{\mg}{\texttt{MadGraph5\_aMC@NLO}}

\begin{document}

\title{HEFT's appraisal of triple (versus double) Higgs weak boson fusion}

\begin{abstract}
Multi-Higgs boson interactions with massive gauge bosons are known to be tell-tale probes of the vacuum manifold of electroweak symmetry breaking. Phenomenologically, a precise determination of these parameters is hampered through increasingly rare processes at the presently available energy frontier provided by the Large Hadron Collider. Contact interactions of three Higgs bosons with the $W$ and $Z$ bosons seem currently well out of experimental reach due to an irrelevant SM production cross section. From a theoretical perspective, in perturbative extensions of the SM such interactions are suppressed by weak loops and further diluted in a priori sensitive processes like weak boson fusion (WBF) when they admit a dimension-six Standard Model Effective Field Theory description. In this work, we identify scenarios that can indeed lead to large, and perhaps even observable modifications of WBF {\emph{triple}} Higgs production most directly parametrised by Higgs Effective Field Theory. We critically analyse these enhancements at the LHC and future colliders from the perspective of unitarity and demonstrate the radiative stability of such analyses under QCD corrections at hadron colliders. Taking into account the restrictions from unitarity, we finally study the expected sensitivity to the electroweak triple Higgs production within HEFT, considering  $HHVV$ and $HHHVV$ effective couplings, at both future hadron and lepton colliders. Particularly, we present numerical predictions for LHC, FCC, CLIC and muon colliders.
\end{abstract}
\author{Anisha\orcidlink{0000-0002-5294-3786}}\email{anisha@glasgow.ac.uk}
\affiliation{School of Physics and Astronomy, University of Glasgow, Glasgow G12 8QQ, United Kingdom\\[0.2cm]}
\author{Daniel~Domenech\orcidlink{0000-0001-5967-9044}}\email{daniel.domenech@uam.es}
\affiliation{Departamento de F\'isica Te\'orica and Instituto de F\'isica Te\'orica, IFT-UAM/CSIC, Universidad Aut\'onoma de Madrid, Cantoblanco, 28049 Madrid, Spain\\[0.2cm]}
\author{Christoph~Englert\orcidlink{0000-0003-2201-0667}}\email{christoph.englert@glasgow.ac.uk}
\affiliation{School of Physics and Astronomy, University of Glasgow, Glasgow G12 8QQ, United Kingdom\\[0.2cm]}
\author{\\Maria~J.~Herrero\orcidlink{0000-0002-2322-1629}}\email{maria.herrero@uam.es}
\affiliation{Departamento de F\'isica Te\'orica and Instituto de F\'isica Te\'orica, IFT-UAM/CSIC, Universidad Aut\'onoma de Madrid, Cantoblanco, 28049 Madrid, Spain\\[0.2cm]}
\author{Roberto~A.~Morales\orcidlink{0000-0002-9928-428X}}\email{roberto.morales@fisica.unlp.edu.ar}
\affiliation{IFLP, CONICET - Departamento de F\'isica, Universidad Nacional de La Plata, C.C. 67, 1900 La Plata, Argentina}
\preprint{IFT-UAM/CSIC-24-111}
\maketitle
\section{Introduction}
\label{sec:intro}
The search for new physics beyond the Standard Model (BSM) has yet to reveal conclusive evidence for the Standard Model's (SM's) ultraviolet completion. In parallel, a relatively detailed picture of the Higgs boson's interactions with known matter has emerged experimentally~\cite{ATLAS:2024fkg,CMS:2024bua}. Furthermore, the increasing data sets of the ATLAS and CMS experiments are now enabling searches and analyses of comparably rare multi-Higgs final states~\cite{ATLAS:2023qzf,CMS:2022hgz}, which provide a more detailed perspective on the electroweak symmetry-breaking sector and fermion mass generation.

So far, all the gathered results strongly suggest that the observations of the Higgs boson are indeed well-described by the SM paradigm. The single Higgs observations so far largely support this view, whereas limits from multi-Higgs production are still relatively far away from the SM in terms of sensitivity. This leaves space for a plethora of phenomenological surprises in the future, chiefly at the high-luminosity (HL) LHC frontier, but also at future experiments superseding the LHC programme.

This work will predominantly focus on the weak bosonic sector of the so-called Higgs Effective Field Theory (HEFT), which is given by
\bear
\label{eq:lag}
\mL_2 &=& \frac{1}{2} \partial_{\mu} H \partial^{\mu} H - \frac12 \mh^2 H^2  \\
&& -\left( \frac12 \kappa_3 \frac{\mh^2}{v} H^3 + \frac{1}{8}\kappa_4 \frac{\mh^2}{v^2} H^4
 \right) \nn \\
&&+\frac{v^2}{4}  F(H) \,\text{Tr}[D_{\mu} U^{\dagger} D^{\mu} U]\nn\\
&&- \frac{1}{2 g^2} \text{Tr}[\hat{W}_{\mu \nu} \hat{W}^{\mu \nu}] - \frac{1}{2 g'^2} \text{Tr}[\hat{B}_{\mu \nu} \hat{B}^{\mu \nu}]\,, \nn  
\eear
in a systematic expansion to leading order momentum $\mO(p^2)$ of a CCWZ~\cite{Coleman:1969sm,Callan:1969sn} $SU(2)_L \times SU(2)_R\to SU(2)_{L+R}\supset U(1)_Y$ parametrisation of the SM gauge sector. The Higgs field $H$ in this formulation takes the form of a custodial singlet reminiscent of a $\sigma/f_0(500)$ meson and we can decorrelate different Higgs-SM matter couplings through the polynomial function $F(H)$. The Goldstone bosons are parametrised as $U=\exp(i\pi^a t^a/v)$ with the would-be Nambu-Goldstone fields $\pi^a$; the $t^a$ label the broken, axial $SU(2)$ generators. The field strengths 
are defined as usual
\begin{equation}
\begin{split}
\hat{W}_{\mu \nu} &= \partial_\mu \hat W_\nu  - \partial_\nu \hat W_\mu + i [ \hat W_\mu,  \hat W_\nu]\,,\\
\hat{B}_{\mu \nu} &= \partial_\mu \hat B_\nu  - \partial_\nu \hat B_\mu \,,~\text{with}\\
 \hat W_\mu  &= g W^a_\mu t^a\,,~
 \hat B_\mu  = g' B_\mu t^3\,,
 \end{split}
 \end{equation}
with weak and hypercharge couplings $g,g'$. Gauge interactions are introduced via the covariant derivative 
\begin{equation}
D_{\mu}U=\partial_{\mu}U+i  \hat{W}_{\mu}U-i U \hat B_{\mu}\,.
\end{equation}
The derivative coupling of the (would-be) Goldstone bosons to the electroweak gauge fields signalises electroweak symmetry breaking; the HEFT parametrisation, as any CCWZ construction, is manifestly concerned with the broken phase of the theory where the $W$ and $Z$ boson have obtained their observed masses. The effective multi-Higgs gauge boson interactions are then given by
\begin{multline}
\label{eq:lag2}
\mL_2 \supset \left(1 + 2 a \frac{H}{v} + b \frac{H^2}{v^2} + c \frac{H^3}{v^3}\right)\\ \times \left(\mw^2 W^+_\mu W^{-\mu} +  {\mw^2 \over 2 \cosw^2} Z_\mu Z^\mu \right)\,,
\end{multline}
for the phenomenologically relevant parametrisation of the contact interactions of the weak gauge boson with up to three Higgs bosons. (The vacuum expectation value is fixed by the $W$ mass $2\mw=gv$ and Weinberg angle is $\cosw=\mw/\mz$). In the SM, we have for unrenormalised quantities
\begin{equation}
F(H) = \left(1+{H\over v} \right)^2\,,
\end{equation}
so that $a=b=1$ and $c=0$. Furthermore, the Higgs potential in the SM maps onto the HEFT parameters $\kappa_{3}=\kappa_{4}=1$.

The relation of the HEFT with the more widely adopted Standard Model Effective Theory (SMEFT)~\cite{Grzadkowski:2010es} is a relevant question. The measurements of Higgs couplings, e.g.~\cite{CMS:2021nnc}, are compatible with a weak doublet character of the electroweak symmetry-breaking vacuum so far. This consistency is currently limited to single Higgs observations which relate to specific parameter choices of the HEFT Lagrangian, which are, however, not motivated beyond any other parameter choice in Eq.~\eqref{eq:lag}. The study of multi-boson final states~\cite{Anisha:2024ljc} is currently underway~(e.g.~\cite{CMS:2022cpr,ATLAS:2023qzf,Brigljevic:2024vuv}) and projected to reach sensitivity to SM production in the gluon fusion channels~\cite{Cepeda:2019klc}. The LHC programme of the near future will therefore clarify whether the wider HEFT ``swampland'' is indeed preferred over the SMEFT-compatible correlations at the weak scale.\footnote{Both HEFT and SMEFT admit coupling choices compatible with the SM, also beyond leading order; HEFT and SMEFT can be identified. In parallel, only HEFT provides a theoretically rigorous extension of the kappa framework of~\cite{LHCHiggsCrossSectionWorkingGroup:2011wcg}.} Although a detailed comparison of HEFT and SMEFT  is definitely beyond the purpose of the present work, we direct the interested reader to Refs.~\cite{Gomez-Ambrosio:2022qsi,Domenech:2022uud,Delgado:2023ynh} where an explicit comparison of the HEFT/SMEFT predictions is done within the context of multiple Higgs production focussing on WBF which is mostly related to our study. 

\begin{figure}[!t]
\centering
\includegraphics[width=0.48\textwidth]{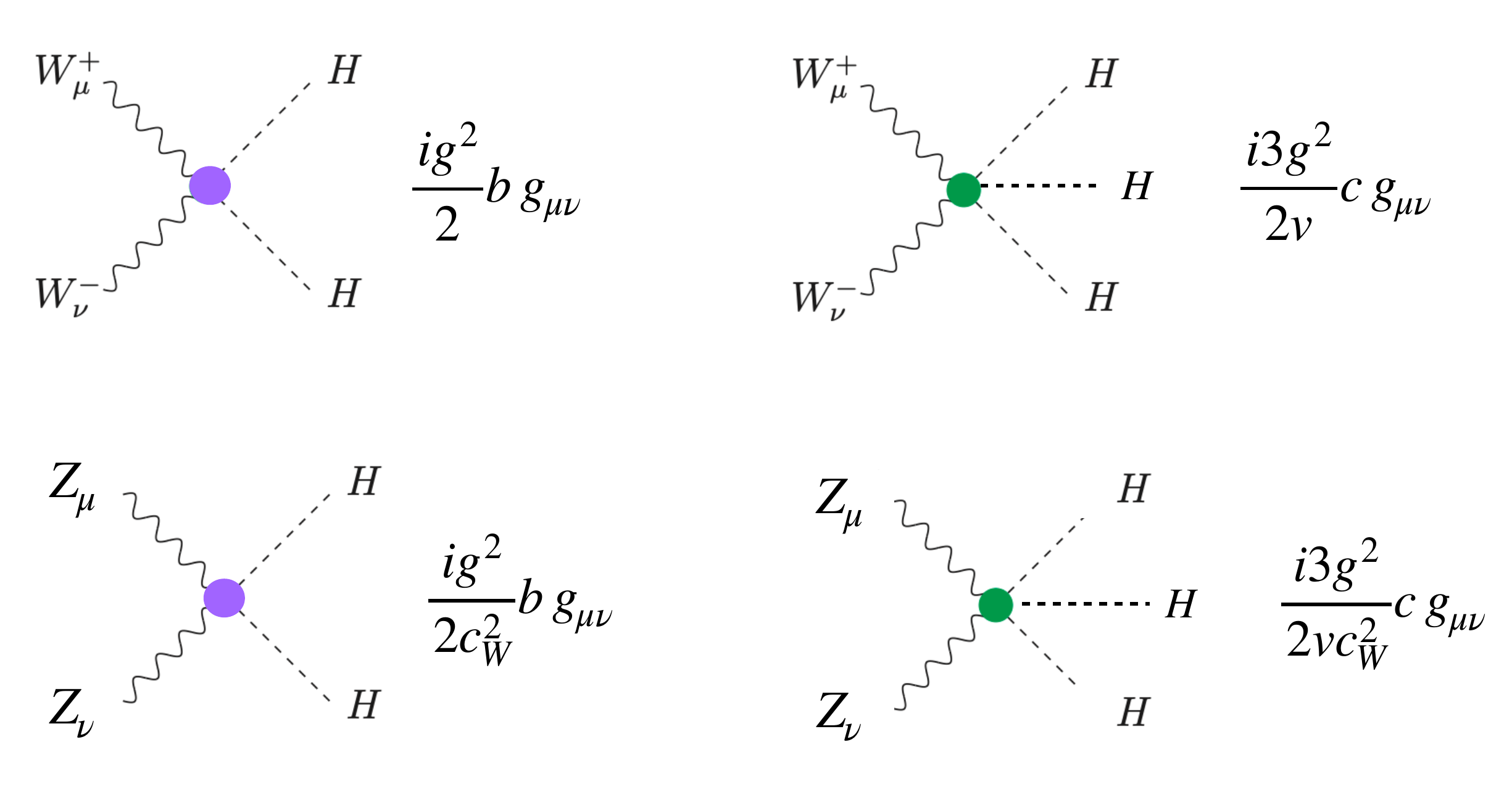}
\caption{Electroweak Feynman rules relevant for $VV\to HH(H)$ scattering probed in WBF multi-Higgs production.\label{fig:frs}}
\vspace{-0.4cm}
\end{figure} 

The HEFT coefficients $a, b,$ and $c$ describe the multi-Higgs contact interactions with the gauge fields, i.e.  $HVV$, $HHVV$ and $HHHVV$ ($V=W,Z$), respectively, see Fig.~\ref{fig:frs}. These parameters are independent due to the singlet nature of the physical Higgs boson in HEFT. In SMEFT these are fully correlated due $SU(2)_L$ gauge symmetry. Operators with a higher mass dimension than four such as the $HHHVV$ contact interactions are loop-suppressed in renormalisable theories such as SM. Beyond dimension 4, at dimension 6 in the SMEFT, only $HHHZZ$ interactions arise from the operator $O_{HD}=(\Phi^{\dagger }D_{\mu }\Phi )^*(\Phi^{\dagger }D^{\mu }\Phi )$ in the Warsaw basis~\cite{Grzadkowski:2010es}, where $\Phi$ denotes SM Higgs doublet. This operator is tightly constrained by measurements of the $T$ parameter (as can be seen from replacing the Higgs legs with their vacuum expectation values in the irreducible $ZZ\to HHH$ diagrams). $HHHWW$ contact interactions arise at dimension-8 level in SMEFT. In the language of Ref.~\cite{Murphy:2020rsh}, these contact interactions are generated mainly from the bosonic class $\Phi^6 D^2$. Including operators from other classes such as $\Phi^4 D^4$, 
and involving field strength tensors in $X^2 \Phi^4$, $X \Phi^4 D^2$ interactions, such contact terms are also generated with novel, non-SM Lorentz structures and momentum dependencies as required by $SU(2)_L$ invariance. 

\begin{figure}[!b]
\includegraphics[width=0.44\textwidth]{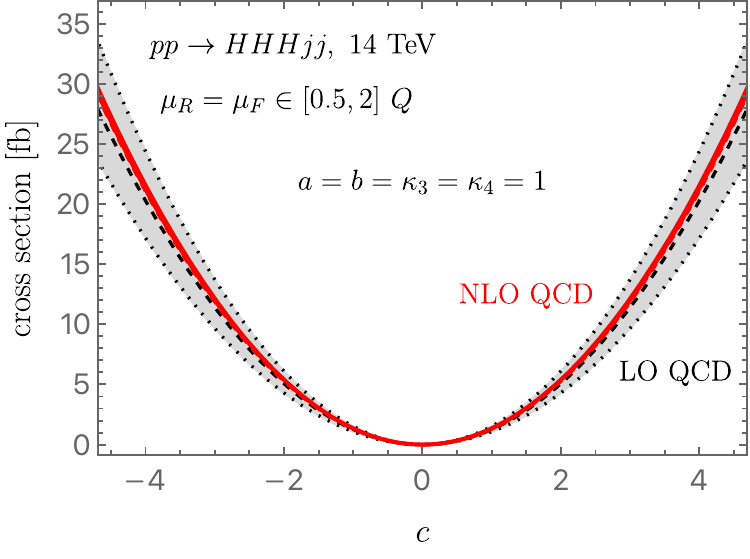}
\caption{\label{fig:xsec} Cross section of $pp \to HHHjj$ for $\mu_R=\mu_F=Q$, i.e. the t-channel momentum transfer of the fermion legs. As can be seen the NLO corrections are modest and well-approximated by this scale choice, eventually rendering the QCD uncertainty negligible. For further details, see the text.}
\vspace{-0.35cm}
\end{figure}

From a technical point-of-view, HEFT can offer some advantages over SMEFT calculations as detailed in~\cite{Herrero:2020dtv,Herrero:2021iqt,Herrero:2022krh}, however, with an opaque power counting (in particular because in the formulation of Eq.~\eqref{eq:lag} a priori different scales are identified with the electroweak vacuum expectation values). This carries the benefit of potentially capturing BSM correlations at intermediate scales towards the SM's UV completion more directly. For instance, theories of strong electroweak symmetry breaking such as the Minimal Composite Higgs Model based on a coset $SO(5)/SO(4)$~\cite{Agashe:2004rs} (MCHM, potentially UV completed in a less minimal scenario~\cite{Ferretti:2014qta,Erdmenger:2020lvq}) directly predict for the effective Higgs interactions in Eq.~\eqref{eq:lag}
\begin{subequations}
\begin{equation}
F(H)={f^2\over v^2} \sin^2\left( {\langle H \rangle + H \over f}\right) \,,
\end{equation}
such that $v=f\sin( \langle H \rangle /f)$ (see also~\cite{Alonso:2016btr}), which leads to
\begin{equation}
\begin{split}
a&=\sqrt{1-\xi} \,,\\
b&=1-2\xi = 2a^2-1\,,\\
c& =-{4\over 3}\xi \sqrt{1-\xi}  = -{4\over 3} a (1-a^2)\,,
\end{split}
\end{equation}
where $\xi=v^2/f^2$ is the electroweak vacuum expectation value $v\simeq 246~\text{GeV}$, sourced by radiative effects, in units of the MCHM pion decay constant $f$. Such a theory is estimated to have an EFT cut-off of $\Lambda \simeq 4\pi f$~\cite{Manohar:1983md} where new strongly coupled resonances are expected enter the picture.\footnote{Just like in QCD, lattice investigations to determine form factors and low-energy constants are necessary~\cite{DelDebbio:2017ini}.} On the other hand, the correlation of $a,b,c,$ in MCHM5 is a manifestation of a $SO(4)\supset SU(2)_L\times U(1)_Y$ fixed point of the model, which enables us to phrase the dynamics as a SMEFT through a field redefinition~\cite{Coleman:1969sm,Alonso:2016oah,Cohen:2020xca} (linked, e.g, to the above $O_{HD}$).\footnote{In compositeness scenarios that go beyond realising the Higgs boson as a pseudo-Nambu Goldstone state, such correlations can be widened~\cite{Sannino:2015yxa}.} Furthermore, specific to MCHM5~\cite{Contino:2006qr} (the functional form of the Higgs potential depends on the particular realisation of partial compositeness), we have (see also~\cite{Grober:2010yv,Gillioz:2012se})
\begin{equation}
\begin{split}
\kappa_3&= {1-2\xi\over \sqrt{1-\xi}} = 2a - {1\over a} \,,\\
\kappa_4&= {1\over 1-\xi} - {28\xi \over 3 } = {28\over 3} (a^2-1) + {1\over a^2} \,.
\end{split}
\end{equation}
\end{subequations}
Another interesting example where the low energy effective couplings of the Higgs boson to the EW bosons and of the Higgs self-couplings could be large is the two Higgs doublet model (2HDM) under specific conditions. It has been shown in Ref.~\cite{Arco:2023sac} that when considering in the 2HDM the heavy mass limit for the BSM Higgs bosons, $H^{\pm}$, $H$ and $A$ ($m_{Heavy}\gg m_h$, where $m_h$ is the light Higgs mass that is identified with the SM-like Higgs), the proper low energy theory for the light modes $h$ and the EW gauge bosons is the HEFT. It turns out that these heavy modes $H^{\pm}$, $H$ and $A$ do not fully decouple in the low energy theory, and their effects appear in the low energy effective Higgs interactions not being suppressed by inverse powers of the heavy masses. This happens, for instance,  in the values found in Ref.~\cite{Arco:2023sac} for $a$, $b$, $\kappa_3$ and $\kappa_4$ which depart from their SM values by corrections that are ${\cal O}(1/m_{Heavy}^0)$  outside the alignment limit (see Eq. (5.12) and Fig. 12 in this reference for the specific values found). In that case, the size of these effective couplings could not be as small as expected.  If instead, the SMEFT is assumed to be the low energy EFT of the 2HDM all the effects from the heavy modes go as inverse powers of the heavy masses (or equivalently, inverse powers of the UV cut-off $\Lambda$) and therefore the size of the effective couplings will be much smaller (Ref.~\cite{Dawson:2023ebe}).

Experimentally, partial insights into the generic HEFT coupling structure have already been obtained. The current results constrain $a$ around unity at the 10\% level~\cite{CMS:2021nnc}. The limits on $b$ are considerably weaker, e.g.,~CMS report $b \in [-0.1 , 2.2 ]$ at 95\% confidence level (CL) in~\cite{CMS:2022cpr} (for a comparable ATLAS analysis see~\cite{ATLAS:2023qzf}); CMS have also constrained $\kappa_3 \in [-0.82,2.94]$ at 95\% CL. No constraint on $c$ has been established at the LHC so far, but the dependence on $\kappa_4$ is known to be shallow~\cite{Stylianou:2023xit,Anisha:2024ljc}; indirect constraints, i.e. $a=a(b,c)$ will unlikely provide significant competitive sensitivity to new physics at the LHC~\cite{Anisha:2022ctm}. High sensitivity to $c,\kappa_4$ can be expected at multi-TeV muon colliders that act as a scattering facility for massive gauge bosons~\cite{Chiesa:2020awd,Han:2021lnp,Celada:2023oji}. The sensitivity to the HEFT parameters has also been explored at future $e^+e^-$ colliders via double and triple Higgs WBF production, see Refs.~\cite{Gonzalez-Lopez:2020lpd,Domenech:2022uud} and below.

What phenomenological value does the search for these contact interactions add to the LHC Higgs programme? As mentioned, in the SM, up to weak radiative corrections, we expect $c=0$ and at leading order in the SM deformation by higher dimensional operators $H^3 ZZ \sim \Delta T^\text{BSM}$. In weak boson fusion $pp\to HHHjj$ which is a priori sensitive to such interactions, the $ZZ\to HHH$ processes are further diluted due to $g'$ suppression of the involved $Z$-fermion couplings. In a nutshell, when assuming a perturbatively coupled extension of the SM, $c$ should be phenomenologically consistent with zero. This extends to the weak gauging of approximate global symmetries of strongly interacting sectors, where we would therefore aim to achieve $\xi \ll 1$, which means ${\cal{O}}(\xi)\simeq | c| \ll a,b\simeq 1$ in the MCHM. 

This changes rather dramatically when considering modifications to the geometry of the electroweak vacuum manifold, including MCHM. For instance, toroidal~\cite{Alonso:2021rac} or modulus-deformed MCHM manifolds~\cite{Englert:2023uug} admit a wider non-SM range of coupling correlations. For the former case, one can show that $a,b$ can be free parameters that determine
\begin{subequations}
\begin{equation}
\label{eq:c1}
c(a,b)= {a\over 3} \left(a^2-b \right) {{3-2a^2-b}\over a^2-1}\quad(a,b\neq 1)\,.
\end{equation}
As remarked by Alonso and West in~\cite{Alonso:2021rac}, the torus admits a choice of coordinates that locally, but not globally, resembles the Standard Model. A parameter choice $a,b=1$ is consistent with 
\begin{equation}
\label{eq:c2}
c=-{2v\over 3f}~(a,b=1)\,.
\end{equation}
\end{subequations} 
In the parametrisation of~\cite{Alonso:2021rac}, this choice is a specific torus $\rho(H)= v + f \sin(H/f), z(H)=-f\cos(H/f)$. Expanding out these coordinates as part of the Goldstone kinetic term directly gives rise to the above relation. Eqs.~\eqref{eq:c1} and \eqref{eq:c2} show that these manifolds imply a highly HEFT-like character. This can be qualitatively understood from the analysis of~\cite{Alonso:2016oah}, which shows that SMEFT-like dynamics rely on a $O(4)$ invariant fix point of the manifold. The fixed point can be identified with the north pole of a four-sphere, which is the manifold of MCHM5, $S^4\simeq SO(5)/SO(4)$; the $SO(4)$ invariance is the rotation around this symmetry axis. The torus, on the other hand, exhibits the same $SO(4)$ invariance around its symmetry axes, which is not part of the manifold. Hence, the torus is intrinsically HEFT-like.

For the deformations considered in~\cite{Englert:2023uug}, we find $-0.2\lesssim c \lesssim 1$ in the region of $a,b$ compatible with current Higgs (WBF pair) analyses. Both these choices can highlight strong coupling as expressed by unitarity considerations of the HEFT, which will be a focus of Sec.~\ref{sec:unit}.

\begin{figure}[!t]
\centering
\includegraphics[width=0.48\textwidth]{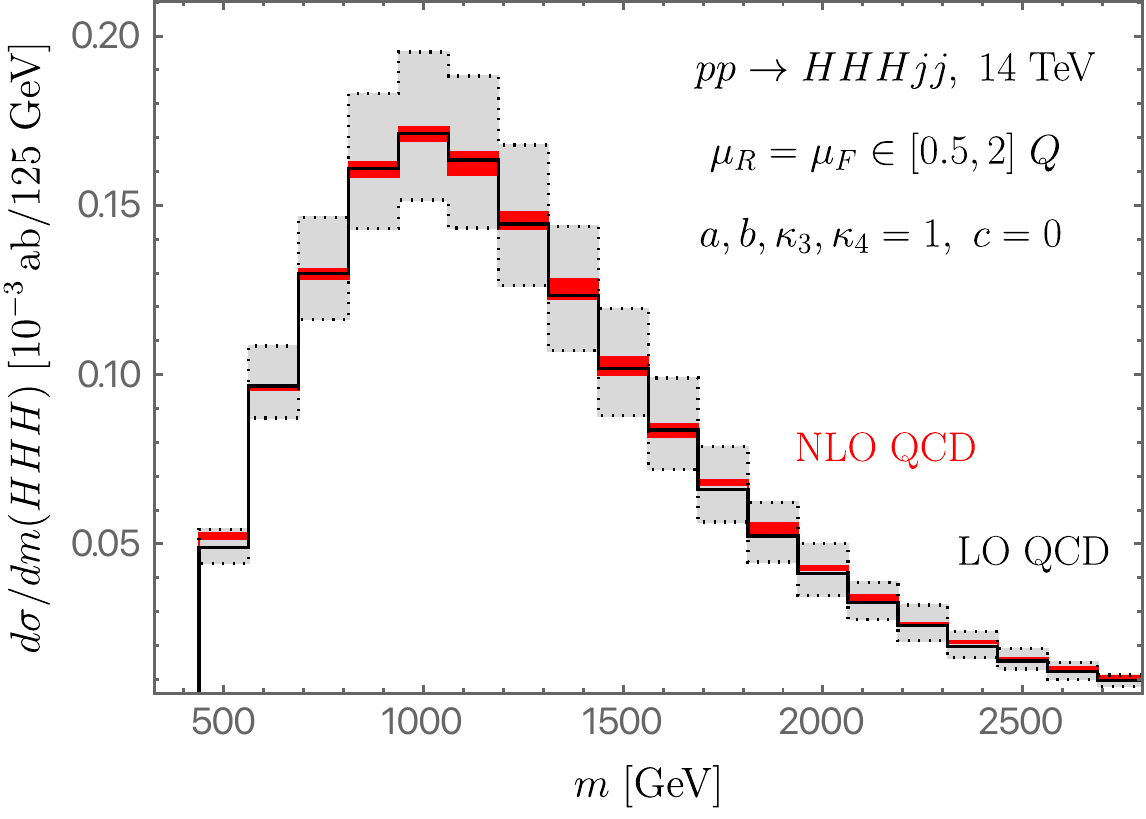}
\caption{\label{fig:diff} Differential invariant triple Higgs mass, $m=m_{HHH}$, in weak boson fusion for the SM coupling choice and scales choice $\mu_R=\mu_F=[0.5,2]~Q$. The central value of the leading order scale improved calculation is included as a solid black line.}
\end{figure}

\begin{figure*}[!t]
	\centering
	\parbox{0.45\textwidth}{\subfigure[\label{fig:HHvsHHHa}]{\includegraphics[width=0.45\textwidth]{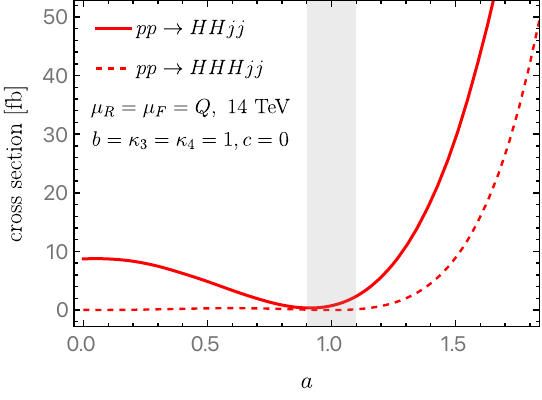}}}\hspace{0.5cm}
	\parbox{0.45\textwidth}{\subfigure[\label{fig:HHvsHHHb}]{\includegraphics[width=0.45\textwidth]{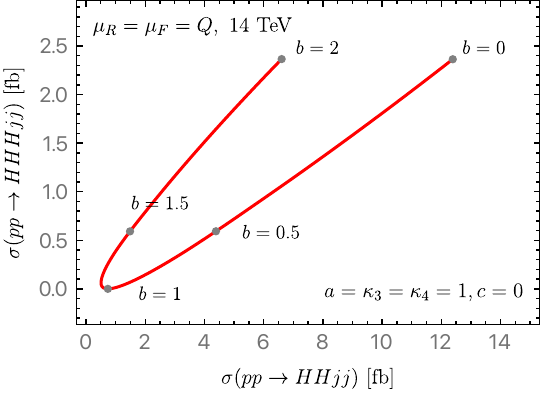}}}\\
	\parbox{0.45\textwidth}{\subfigure[\label{fig:HHvsHHHc}]{\includegraphics[width=0.45\textwidth]{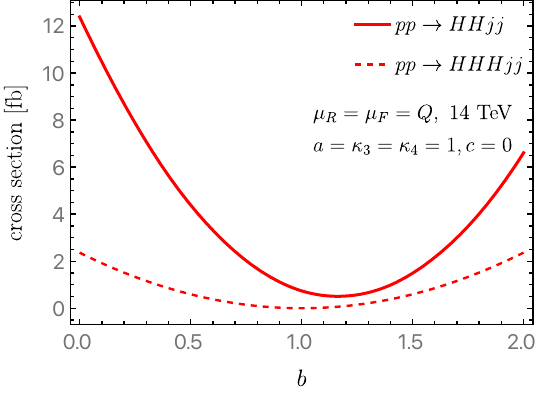}}}\hspace{0.5cm}
	\parbox{0.45\textwidth}{\parbox{0.37\textwidth}{\vspace{1.2cm}\caption{Comparison of $HHjj$ and $HHHjj$ production including NLO QCD corrections discussed in the text. We show a wide range of $a$ in \ref{fig:HHvsHHHa} for illustration purposes. $a$ is already constrained, e.g.,~\cite{CMS:2021nnc} shown as the grey band. \ref{fig:HHvsHHHb} shows the multi-Higgs WBF cross sections correlation as function of $b$ which is further discussed in Sec.~\ref{sec:unit}. A priori, large enhancements are possible in both $HHjj$ and $HHHjjj$ production \ref{fig:HHvsHHHc}.}
	\label{fig:HHvsHHH}}}
	\vspace{-0.3cm}
\end{figure*} 
The examples of the last paragraph show that searches for a sizeable $c$ within the currently allowed single and double Higgs constraints are by no means theoretically futile. They offer a deeper understanding of the geometry of the electroweak vacuum. The expectation of $c\simeq 0$ in standard (and experimentally challenged) BSM scenarios is a strong motivation to work towards perturbatively meaningful constraints of $c\neq 0$ on WBF triple Higgs production (see also~\cite{Belyaev:2018fky,Mahmud:2024iyn}). This will be the focus of the remainder of this work, which is organised as follows.  In Sec.~\ref{sec:lhc}, we will discuss the LHC prospects for the considered processes of triple and double Higgs production from WBF.  We then critically assess the expected cross section enhancements at LHC from the perspective of leading order unitarity (Sec.~\ref{sec:unit}). Finally,  in  Sec.~\ref{sec:future}  and by taking also into account the restrictions from unitarity, we study the expected sensitivity to the electroweak triple Higgs production within HEFT, considering  $HHVV$ and $HHHVV$ effective couplings,  at both future hadron and lepton colliders.  Particularly, we present numerical predictions for LHC, FCC, CLIC and muon colliders.  Our conclusions are outlined in Sec.\ref{sec:conc}.
\vspace{-0.2cm}
\section{LHC predictions}
\label{sec:lhc}
Weak boson fusion and its phenomenological appeal have been discussed in the existing literature~\cite{Cahn:1983ip,Han:1992hr} including precision predictions~\cite{Berger:2004pca,Figy:2003nv,Figy:2004pt,Ciccolini:2007jr,Cacciari:2015jma}. The phenomenological signature of weak boson fusion is highly distinct; the production of a heavy QCD-singlet via the $t$-channel exchange results in two energetic forward jets at moderate transverse momentum with weak decay products localised in the central part of the detector. Jet vetos~\cite{Barger:1994zq} then act as additional handles to remove irreducible contamination from gluon-fusion contributions, which is not a desirable avenue for multi-Higgs production that relies on the dominant decay modes $h\to b\bar b$. Different methods have been discussed~\cite{Andersen:2012kn,Dolan:2015zja}, and are employed~\cite{ATLAS:2023qzf,CMS:2022hgz} to obtain high-purity multi-Higgs WBF selections. However, it is clear that gaining experimental sensitivity to the more elusive triple Higgs WBF processes poses a formidable challenge along these lines as well, and any successful strategy will necessarily need to be based on a proliferation of multivariate and machine learning techniques to optimally exploit correlations when facing up overwhelming backgrounds.

Whilst it is beyond the scope of this work to estimate the experimental sensitivity (or lack thereof) that can be gained to this final state at the fully-showered and hadronised stage, we can argue for the reward that such an analysis can hold. In Fig.~\ref{fig:xsec}, we show the $pp\to HHHjj$ cross sections as a function of $c$ for $a=b=\kappa_3=\kappa_4=1$ for a wide range of $c$. As we will discuss below (and also see~\cite{Celada:2023oji}) some of this parameter region is constrained by electroweak unitarity constraints, however, as for single and double Higgs production~\cite{Figy:2008zd}, the QCD corrections to WBF triple Higgs production are under excellent perturbative control. This is further highlighted (for single and double Higgs production) through the possibility of adapted scale choices that re-sum a large share of the (relatively moderate) QCD corrections as part of QCD-factorisation. More concretely, when the relevant factorisation and renormalisation scales are chosen to be the $t$-channel momentum transfer of the corresponding quark leg, the leading order result approximates the next-to-leading order result very well, with scale uncertainties being reduced to below 1\% for typical WBF cuts
\begin{equation}
\label{eq:cuts}
\begin{split}
&p_{T,j}\geq 20~\text{GeV},~m_{jj}\geq 500~\text{GeV}\,,\\
&2<|\eta_j|<5,~\eta_{j_1}\times \eta_{j_2}<0\,,
\end{split}
\end{equation}
for the two tagging jets of the WBF event at a 14 TeV LHC (i.e. forward jets in different detector hemispheres at moderate transverse momentum with large invariant mass). We note that in both our generated events and numerical predictions in the present work,  and for all colliders considered,  we have accounted for all relevant configurations contributing to electroweak triple and double Higgs production.  This includes all contributing diagrams, not only those mediated by WBF subprocesses. For instance, in double Higgs production in $pp$ collisions, we consider all diagrams for $q_1 q_2 \to HH q_3 q_4$  (with $q_i$ refereeing generically to both quarks and anti-quarks). These diagrams include those with WBF topology, like $q_1 q_2 \to W^*W^* q_3 q_4 \to HH q_3 q_4$ as well as other topologies, like Higgs-strahlung diagrams,  $q_1q_2 \to H^*Z^*\to HH q_3 q_4$, etc.  We have also checked that, as expected, after applying the optimised WBF cuts to the final $q$-jets, we select quite efficiently the dominant WBF events.  Similar comments apply to the case of lepton colliders where,  again,  all configurations contributing to $l^+ l^- \to HH\nu \bar \nu$ and $l^+l^- \to HHH \nu \bar \nu$ have been considered in our work.  In that case,  the cuts for the WBF selection are applied to the missing transverse energy carried by the final neutrino-antineutrino pairs.

 For the SM, $a,b,\kappa_3,\kappa_4=1,c=0$, we find a (phenomenologically hardly relevant) WBF cross section at LHC for $\mu_R=\mu_F=Q$\footnote{The hadron collider calculations performed in this work are available as an add-on to the recent {\tt{vbfnlo3}} release~\cite{Baglio:2024gyp,Arnold:2008rz} as part of this paper. The relevance of the WBF topologies compared to negligible associated Higgs production processes for the chosen cuts has been checked explicitly against \mg~\cite{Alwall:2014hca}. The inclusion and consideration of the five-particles vertices in Fig.~\ref{fig:frs} in {\tt{vbfnlo3}} in the context of scale-adapted QCD corrections is new.  Users can simultaneously modify any electroweak coupling detailed in this work and obtain QCD-corrected differential cross sections straightforwardly. The comparison against \mg~was facilitated using \texttt{FeynRules}~\cite{Alloul:2013bka} and \texttt{UFO}~\cite{Alloul:2013bka}.}
\begin{equation}
\sigma_\text{NLO}(pp\to HHH j j) = 0.197~\text{ab}
\end{equation}
with a $K=\sigma_\text{NLO}/\sigma_\text{LO}$ factor well below 1\%. Hence, the $pp \to HHH jj$ provides a clean window into studying weak contact interactions unique to the $H^3$ process. The relevant differential distribution of the invariant triple Higgs mass is also stable with a flat differential $K$ factor, Fig.~\ref{fig:diff}. This behaviour extends to non-SM coupling choices as well.

\begin{figure}[!t]
\includegraphics[width=0.45\textwidth]{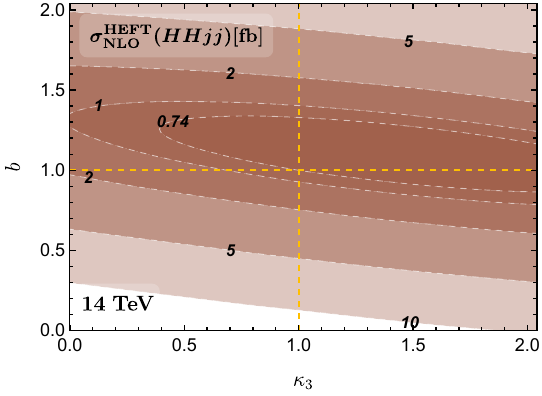}
\caption{\label{fig:2h}Weak boson fusion di-Higgs production as a function of $b,\kappa_3$ for the cuts detailed in the text, including NLO QCD corrections for a scale choice $\mu_R=\mu_F=Q$.}
\label{fig-k3_b_HH}
\end{figure}

For a comparison of $HHjj$ and $HHHjj$ production at the LHC (for $\sqrt{s}=14$ GeV), we show in Fig.~\ref{fig:HHvsHHH} the HEFT cross section predictions as a function of the parameters $a$ and $b$ for $\kappa_3=\kappa_4=1, c=0$ and the cuts specified in Eq.~\eqref{eq:cuts}; contours of the relevant parameters $b,\kappa_3$ after constraints on $a\simeq 1$ are taken into account are shown in Fig.~\ref{fig:2h}. The WBF topologies to both the $HHjj$ and $HHHjj$ processes dominate the rates for WBF cuts (e.g.~Eq.~\eqref{eq:cuts}, see also~\cite{Ciccolini:2007jr}), which we have checked against a comparison with~\mg. Based on the results of~\cite{Bredenstein:2008tm}, we also expect QCD-electroweak interference to be negligible in this phase space region. Note that the dependence on $a$ for both processes is heavily constrained by LHC single Higgs observations (with $a_{\rm max} \sim 1.13$). Nonetheless, we observe a high sensitivity in both processes to $b$, for the interval currently allowed by data. There is a clear correlation between the $HHjj$ and $HHHjj$ rates, which is manifest in Fig.~\ref{fig:HHvsHHHb} where the cross sections are displayed in the vertical/horizontal axis for values in the currently allowed interval $b \simeq [0,2]$. The high-energy behaviour of the amplitudes is given by
\begin{equation}
\label{eq:contour}
\begin{split}
 {\cal{M}} (W_L^+W_L^- \to HH )& \sim g^2( b-a^2 ) { s \over 4 \mw^2} \,,\\
 {\cal{M}} (W_L^+W_L^- \to HHH )& \sim {3g^3 \over 2\mw} \left(c - {4\over 3} a(b-a^2)\right) {s\over 4\mw^2}\,.
\end{split}
\end{equation}
for $s\gg \mw^2,\mh^2$ and longitudinal polarisations (in this region the effective $W$ approximation is motivated \cite{Dawson:1984gx}). The high-energy limit for $HHH$ production has been provided in Ref.~\cite{Delgado:2023ynh}, relying on the Goldstone equivalence theorem. Similar equations hold for $ZZ$ due to the custodial symmetric coupling structures. This shows that the $a-b^2$ interplay probed in $pp\to HHjj$ also leaves a phenomenology-shaping imprint in the $HHHjj$; our cross section contours (and exclusions) reflect the above unitarity relations even beyond the high-energy approximation that underpins the equivalence theorem (see below).

\begin{figure*}[!t]
\centering
\parbox{0.45\textwidth}{\subfigure[\label{fig:3hca}]{\includegraphics[width=0.45\textwidth]{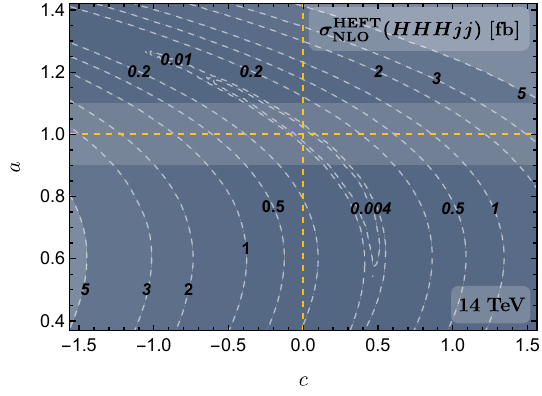}}}\hspace{0.5cm}
\parbox{0.45\textwidth}{\subfigure[\label{fig:3hcb}]{\includegraphics[width=0.45\textwidth]{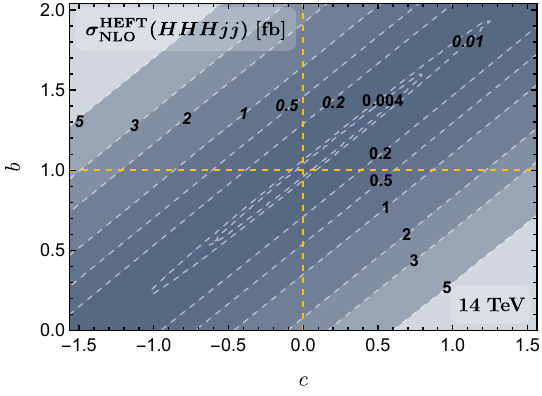}}}\\
\parbox{0.45\textwidth}{\subfigure[\label{fig:3hck3}]{\includegraphics[width=0.45\textwidth]{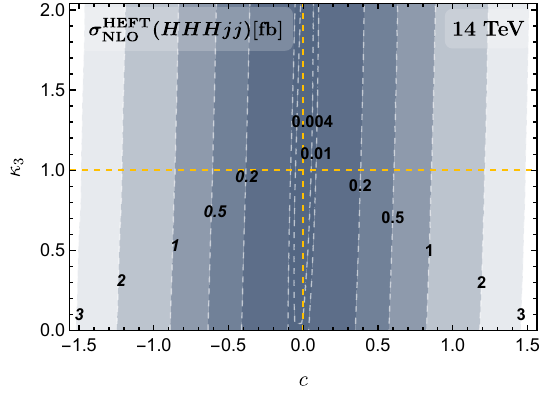}}}\hspace{0.5cm}
\parbox{0.45\textwidth}{\subfigure[\label{fig:3hck4}]{\includegraphics[width=0.45\textwidth]{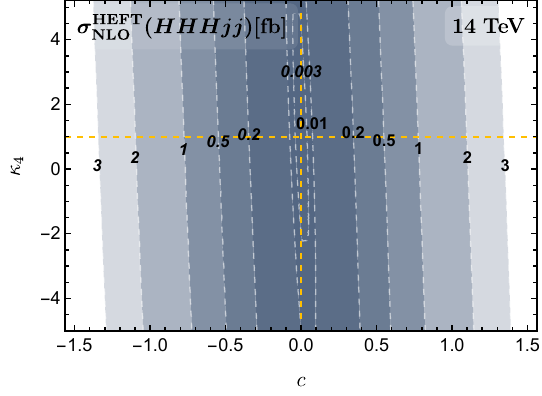}}}
\caption{\label{fig:3h}NLO QCD cross section contours for WBF triple Higgs production for the selection cuts of Eq.~\eqref{eq:cuts} as functions of (a) $c,a$ (including single Higgs constraints on $a$), (b) $c,b$, (c) $c,\kappa_3$, and (d) $c,\kappa_4$. Again scale choices are $\mu_R=\mu_F=Q$. Couplings that are not scanned over are fixed to their would-be SM values $a,\kappa_3,\kappa_4=1$.}
\vspace{-0.3cm}
\label{fig-NLOQCD}
\end{figure*}

In Figs.~\ref{fig:2h} and~\ref{fig:3h}, we show cross section contours for the different HEFT parameter choices that affect $pp\to HH(H)jj$ production. Di-Higgs WBF production serves to set limits on $a,b$, and to a lesser extent $\kappa_3$ as done by the experiments already. Triple Higgs production opens up the a priori possibility to probe $c$ and $\kappa_4$. The findings of gluon fusion triple Higgs production regarding the relative insensitivity to $\kappa_4$~\cite{Stylianou:2023xit,Anisha:2024ljc} are mirrored by WBF production. Again the unitarity relations of Eq.~\eqref{eq:contour} are probed in these processes, which also explains ellipsoid-shaped cross section contours of $\sigma(a,c)$  Fig.~\ref{fig:3hca}, the linear dependence of $\sigma(b,c)$ in Fig.~\ref{fig:3hcb} and the relative insensitivity to $\kappa_{3,4}$.  As unitarity underpins the phenomenology of the potential enhancements of the rates of the considered processes, we focus on the robustness of limits in the next section. In Figs.~\ref{fig-k3_b_HH} and~\ref{fig-NLOQCD}, we show two-dimensional projections of the sensitivity to two  HEFT effective couplings,  $a$,  $b$ and $c$ (couplings of Higgs bosons to EW gauge bosons) on the one hand,  and $\kappa_3$ and $\kappa_4$ (triple and quartic Higgs self-couplings) on the other hand,  for illustration purposes.  It is apparent that the sensitivity to the first set of couplings is higher in WBF. Therefore, in the following, we will focus on the protagonist HEFT couplings $a$,  $b$ and $c$.

\section{Unitarity constraints on $HH(H)$  production from WBF}
\label{sec:unit}
The HEFT predictions of multi-Higgs production from WBF, like  $WW \to HH$ and $WW \to HHH$ violate perturbative unitarity at high energies in the TeV region, Eqs.~\eqref{eq:contour} (see also~\cite{Maltoni:2001dc,Belyaev:2018fky,Alonso:2021rac,Mahmud:2024iyn}). 
As $a\simeq 1$ has been established with recent Higgs measurements unitarity constraints from this parameter are no longer driving overall perturbativity considerations.
The presence of $b \neq 1$, however, implies pertubative unitarity violation in di-Higgs and triple Higgs production; the presence of $c\neq 0$ is linked to unitarity violating cross section enhancement of $HHHjj$ production. The case of di-Higgs production has been studied previously in the literature and the constraints on the values of the $a$ and $b$ HEFT parameters have also been derived (see, for instance, \cite{Gonzalez-Lopez:2020lpd, Anisha:2022ctm}). The unitarity constraint on $\kappa_3$ from di-Higgs and the constraint on $\kappa_4$ from triple Higgs are much milder (e.g.~\cite{Gonzalez-Lopez:2020lpd}) and practically do not affect the LHC multi-Higgs boson rates as they do not drive a kinematic cross section enhancement. Hence, in the following, we focus on the $b$ and $c$ parameters and their limit setting. We set the remaining parameters to their respective SM values, i.e. $a=\kappa_3=\kappa_4=1$.  As quantitatively identical bounds are provided by the $ZZ\to HH(H)$ amplitudes, we focus on the $WW$ processes in this section.

In this study,  for the case of $W_LW_L \to HH$,  we apply the so-called elastic unitarity requirement on the dominant partial wave amplitude, with  $J=0$ 
\begin{equation}
|a_0(W_LW_L \to HH)(s)| \leq 1\,,
\label{unitHH}
\end{equation}
where $\sqrt{s}=m_{HH}$ is the invariant double-Higgs mass. This elastic unitarity condition is less restrictive than the so-called fully inelastic unitarity condition, $|a_0| \leq 1/2$,  which takes into account that the scattering process $WW \to HH$ is indeed an inelastic process in contrast to $WH \to WH$, which is an elastic process (see,  for instance, the discussion in ~\cite{Kilian:2018bhs})\footnote{We wish to thank Antonio Dobado and the anonymous PRD referee for their comments and clarifications on these\\ elastic/inelastic different partial wave conditions.} 
We have checked that the differences in applying the elastic or fully inelastic unitarity conditions have a minor impact on the final numerical results of the present work. This constraint is widely discussed in the literature, but of course, other measurements are also known, e.g.~\cite{Kilian:2018bhs,Stylianou:2023xit}. Unitarity constraints always give qualitative constraints related to perturbative (non-)convergence. The Lagrangian that we consider is manifestly hermitian, implying a unitary time evolution.
Similarly, for the case of $W_LW_L \to HHH$, we apply the unitarity restriction on the inelastic cross section  (see, e.g.,~\cite{Maltoni:2001dc,Belyaev:2018fky,Gonzalez-Lopez:2020lpd,Domenech:2022uud})  
\begin{equation}
\sigma(W_L W_L \to HHH) \leq {4\pi \over s}
\label{unitHHH}
\end{equation}
where  $\sqrt{s}=m_{HHH}$ is the invariant triple-Higgs mass.

\begin{figure}[!t]
\includegraphics[width=0.5\textwidth]{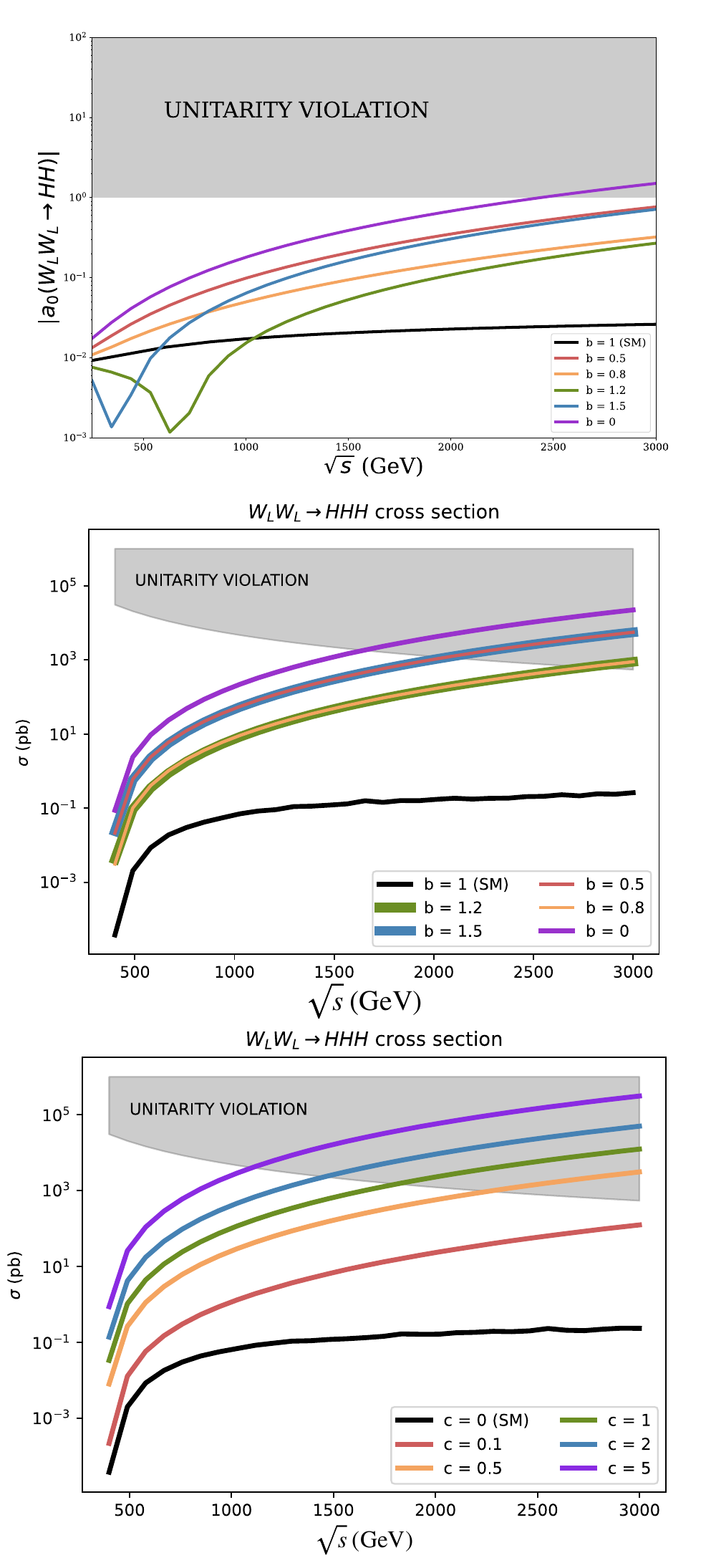}
\caption{Unitarity constraints on double Higgs production (upper plot) as a function of $b$ and on triple Higgs production as a function of  
$b$ (middle plot) and of $c$ (lower plot).}
\label{fig:unitarity}
\end{figure}

The results of these unitarity requirements are displayed in Fig. \ref{fig:unitarity}. The upper plot is for $W_LW_L \to HH$ as a function of $b$, and the middle and lower plots are for $W_LW_L \to HHH$ as a function of $b$ and $c$, respectively. The predictions for the SM choices that all preserve unitarity are also included for comparison. Notice that we show predictions up to $3~\text{TeV} \simeq 4 \pi v$, the typical energy scale controlling the chiral expansion. Comparing the first and second plots, we see that the constraints on $b$ are stronger in triple Higgs than in double Higgs production. For instance, for $b=0$, the crossing into the unitarity violating region occurs at around 2.5 TeV in $HH$, whereas it happens at around 1.65~TeV in $HHH$ production. In fact, for all the $b \neq 1$ values explored in this figure, the $HHH$ cross section enters the unitarity-violating region below 3 TeV. The third plot displays the unitarity crossing energy point for each of the $c \neq 0$ values explored in the case of $WW\to HHH$. For $c=$ 0.5,  1,  2 and 5 this crossing occurs below 3 TeV,  concretely at around 2.25~TeV, 1.75~TeV, 1.4 TeV~and 1.2 TeV, respectively.  The case with $c=0.1$ enters the unitarity violating region above 3 TeV, at around 3.7~TeV.

\begin{figure}[!t]
\includegraphics[width=0.5\textwidth]{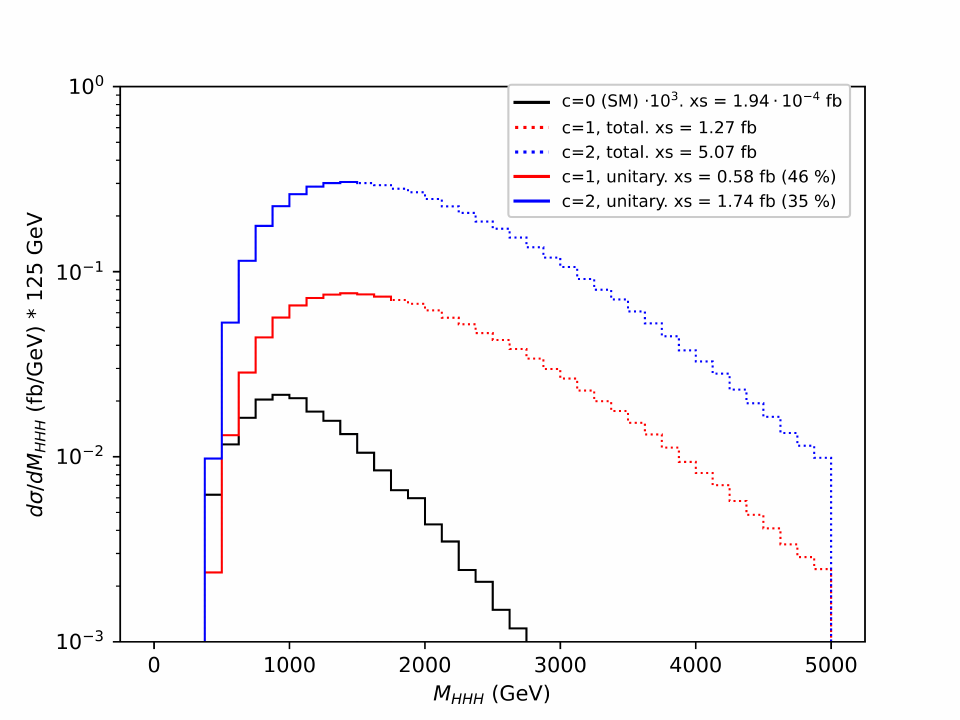}
\caption{Unitarity constraints on the distributions with the triple Higgs invariant mass $m_{HHH}$ for two examples with $c=1$ (in red) and $c=2$ (in blue).  Solid lines are predictions preserving unitarity.  Dashed lines are predictions that do not preserve unitarity. The SM is included (scaled by $10^3$) for comparison as a black line. Here we use LO QCD given that the overall corrections are small.}
\label{fig:unitaritydist}
\end{figure}

The issue of unitarity violation in triple Higgs production from WBF within the context of an effective Lagrangian approach with similar interactions as ours in Eq. \eqref{eq:lag}, but when restricting them to the unitary gauge,  have been considered previously by Kilian et al. in Ref.~\cite{Kilian:2018bhs}. The relations between the relevant effective couplings in this reference and ours can be found by comparing our Eq. \eqref{eq:lag2} with their Eq. (2.4). We find the following relations: $a=g_{W,a1}=g_{Z,a1}$, $b=g_{W,a2}=g_{Z,a2}$ and $c=g_{W,a3}=g_{Z,a3}$. Thus, whereas in our approach there are just three independent effective couplings, $a$,$b$ and $c$ for one, two and three Higgs bosons to electroweak gauge bosons, respectively, in their approach they have six independent, three for the couplings to $W$ bosons, $g_{W,a1,a2,a3}$, and three for the couplings to $Z$ bosons,  $g_{Z,a1,a2,a3}$.  Reference ~\cite{Kilian:2018bhs} presents a detailed study of unitarity constraints including analytic restrictions on partial wave amplitudes.  In Fig. \ref{fig:unitarity}, we  explore the unitarity constraints numerically as a function of the subprocess energy. The most important outcome from this comparison is that, for the energy values of $M_{HHH}$ of order a few TeV, the electroweak triple Higgs production at future colliders will be reduced by important constraints on the size of the $b$ and $c$ effective couplings. We have checked that our constraints from Fig. \ref{fig:unitarity} are in qualitative agreement with the constraints in Ref.~\cite{Kilian:2018bhs}.

The perturbative consistency constraints are only relevant if these regions are statistically explored by the collider experiments and used to derive limits. This is, in particular, relevant for future colliders, which are envisioned to target partonic collision energies in the range of 10~TeV. For the LHC, we show in Fig. ~\ref{fig:unitaritydist} the predicted cross section distributions with $m_{HHH}$ within the HEFT for two BSM examples with $c=1$ and $c=2$.  The solid lines refer to the predictions in the unitarity-preserving region and the dashed lines extend into the unitarity-violating phase space region. The crossing points are at the above commented values of  $m_{HHH}=1.75~\text{TeV}$ and 1.4 TeV, respectively. One can then estimate the restrictions imposed from unitarity constraints on the LHC rates by computing a reduction factor $\epsilon_U$ which is defined as the following ratio
\begin{equation}
\epsilon_U \equiv \frac{\rm rates \,  preserving\, unitarity}{\rm total \, rates}\,,
\label{epsilonU} 
\end{equation}
an experimental analysis should rely on $\epsilon_U \sim 1$ as it will most likely be based on a non-differential cross section exclusion. Not considering selection efficiencies or backgrounds, we find for the LHC ($\sqrt{14}$ GeV): $\epsilon_U=0.46$ for $c=1$ and  $\epsilon_U=0.35$ for $c=2$ (a more comprehensive overview is given in Tab.~\ref{tab:lhc}). A region $|c| < 0.5$, if experimentally accessible, would be perturbatively controlled, whilst providing insights into the electroweak vacuum as outlined in Sec.~\ref{sec:intro}.

\sisetup{
  scientific-notation = true,
  table-number-alignment = center
}

\begin{table*}[p]
\renewcommand{\arraystretch}{1.25} 
\begin{center}
\parbox{0.99\textwidth}{
  \begin{minipage}{.1\linewidth}
        \setlength{\tabcolsep}{0.2cm}
    \begin{tabular}{|c|c|}        
        \multicolumn{2}{c}{{}}\\
        \hline
        $b$ &  $c$  \\ \hline
        1.0 & 0.0 \\
        1.0 & 0.1 \\
        1.0 & 0.5 \\
        1.0 & 1.0 \\
        1.0 & 2.0 \\
        1.0 & 5.0 \\
        1.5 & 0.0 \\
        1.5 & 0.1 \\
        1.5 & 0.5 \\
        1.5 & 1.0 \\
        1.5 & 2.0 \\
        1.5 & 5.0 \\
        1.2 & 0.0 \\
        1.2 & 0.1 \\
        1.2 & 0.5 \\
        1.2 & 1.0 \\
        1.2 & 2.0 \\
        1.2 & 5.0 \\
        0.8 & 0.0 \\
        0.8 & 0.1 \\
        0.8 & 0.5 \\
        0.8 & 1.0 \\
        0.8 & 2.0 \\
        0.8 & 5.0 \\
        0.5 & 0.0 \\
        0.5 & 0.1 \\
        0.5 & 0.5 \\
        0.5 & 1.0 \\
        0.5 & 2.0 \\
        0.5 & 5.0 \\
        0.0 & 0.0 \\
        0.0 & 0.1 \\
        0.0 & 0.5 \\
        0.0 & 1.0 \\
        0.0 & 2.0 \\
        0.0 & 5.0 \\
        \hline
    \end{tabular}
    \end{minipage}
    \begin{minipage}{.37\linewidth}
            \setlength{\tabcolsep}{0.3cm}
    \begin{tabular}{|S[table-format=1.3e-1]|S[table-format=1.3e-1]|c|}
        \hline
        \multicolumn{3}{|c|}{{HL-LHC (14 TeV, 3 $\text{ab}^{-1}$)}}\\
        \hline
        {$\sigma$ (fb)} & {$\epsilon_{U}$} &  {$N_{U}^{6b}$} \\ \hline
        1.94e-04 & 1.00e+00 &         0 \\
        1.28e-02 & 9.96e-01 &         2 \\
        3.17e-01 & 6.84e-01 &        33 \\
        1.27e+00 & 4.60e-01 &       89 \\
        5.07e+00 & 3.50e-01 &       272 \\
        3.17e+01 & 1.69e-01 &       821 \\
        5.56e-01 & 6.36e-01 &        54 \\
        4.01e-01 & 6.35e-01 &        39 \\
        3.53e-02 & 9.33e-01 &         5 \\
        1.48e-01 & 7.77e-01 &        18 \\
        2.27e+00 & 4.75e-01 &       166 \\
        2.39e+01 & 1.70e-01 &       623 \\
        8.90e-02 & 8.37e-01 &        11 \\
        3.46e-02 & 9.36e-01 &         5 \\
        7.06e-02 & 8.96e-01 &        10 \\
        6.86e-01 & 5.84e-01 &        61 \\
        3.82e+00 & 4.14e-01 &       242 \\
        2.84e+01 & 1.69e-01 &       737 \\
        8.95e-02 & 8.38e-01 &        12 \\
        1.69e-01 & 7.69e-01 &        20 \\
        7.41e-01 & 5.85e-01 &        67 \\
        2.03e+00 & 4.71e-01 &       147 \\
        6.50e+00 & 2.88e-01 &       287 \\
        3.51e+01 & 1.16e-01 &       624 \\
        5.58e-01 & 6.37e-01 &        54 \\
        7.38e-01 & 5.85e-01 &        66 \\
        1.71e+00 & 4.71e-01 &       124 \\
        3.50e+00 & 4.17e-01 &       224 \\
        8.98e+00 & 2.25e-01 &       310 \\
        4.06e+01 & 1.15e-01 &       720 \\
        2.23e+00 & 4.70e-01 &       161 \\
        2.58e+00 & 4.10e-01 &       162 \\
        4.22e+00 & 2.86e-01 &       185 \\
        6.85e+00 & 2.85e-01 &       300 \\
        1.40e+01 & 2.25e-01 &       484 \\
        5.07e+01 & 1.15e-01 &       892 \\
        \hline
    \end{tabular}
  \end{minipage}
  \begin{minipage}{.4\linewidth}
              \setlength{\tabcolsep}{0.35cm}
    \begin{tabular}{|S[table-format=1.3e-1]|S[table-format=1.3e-1]|c|}
        \hline
        \multicolumn{3}{|c|}{{FCC-hh (100 TeV, 30 $\text{ab}^{-1}$)}}\\
        \hline
        {$\sigma$ (fb)} & {$\epsilon_{U}$} & {$N_{U}^{6b}$} \\ \hline
        1.27e-02 & 1.00e+00 &        19 \\
        1.25e+01 & 1.57e-01 &      3005 \\
        3.12e+02 & 4.91e-02 &     23511 \\
        1.25e+03 & 3.02e-02 &     57887 \\
        4.99e+03 & 1.18e-02 &     90194 \\
        3.12e+04 & 5.68e-03 &    271802 \\
        5.50e+02 & 4.11e-02 &     34653 \\
        3.97e+02 & 4.12e-02 &     25060 \\
        3.41e+01 & 1.14e-01 &      5959 \\
        1.42e+02 & 6.91e-02 &     15074 \\
        2.23e+03 & 2.49e-02 &     85013 \\
        2.35e+04 & 5.83e-03 &    209782 \\
        8.79e+01 & 8.05e-02 &     10867 \\
        3.42e+01 & 1.14e-01 &      5982 \\
        6.88e+01 & 1.01e-01 &     10684 \\
        6.73e+02 & 3.69e-02 &     38136 \\
        3.75e+03 & 1.96e-02 &    112846 \\
        2.80e+04 & 5.91e-03 &    253718 \\
        8.80e+01 & 8.02e-02 &     10824 \\
        1.67e+02 & 6.37e-02 &     16284 \\
        7.31e+02 & 3.65e-02 &     40929 \\
        2.00e+03 & 2.48e-02 &     76125 \\
        6.40e+03 & 1.18e-02 &    116163 \\
        3.46e+04 & 3.71e-03 &    196885 \\
        5.50e+02 & 4.16e-02 &     35126 \\
        7.28e+02 & 3.89e-02 &     43399 \\
        1.69e+03 & 2.47e-02 &     64102 \\
        3.45e+03 & 1.96e-02 &    103723 \\
        8.85e+03 & 8.44e-03 &    114532 \\
        4.00e+04 & 3.84e-03 &    235680 \\
        2.20e+03 & 2.40e-02 &     81140 \\
        2.54e+03 & 1.86e-02 &     72706 \\
        4.17e+03 & 1.13e-02 &     72099 \\
        6.76e+03 & 1.12e-02 &    116597 \\
        1.38e+04 & 8.29e-03 &    175548 \\
        4.99e+04 & 3.84e-03 &    293866 \\
        \hline
    \end{tabular}
  \end{minipage}
}
\caption{HEFT rates predicted with {\tt{vfbnlo}} (and LO QCD) for triple Higgs production at future $pp$ colliders,   HL-LHC (14 TeV, 3 $\text{ab}^{-1}$)(left subtable) and FCC-hh (100 TeV, $30~\text{ab}^{-1}$) (right subtable) for some selected values of $b$ and $c$. The first row displays the SM rates. Here, $\sigma \equiv \sigma(pp  \rightarrow HHHjj$),  $\epsilon_U \equiv {\rm rates\, preserving\, unitarity}/ {\rm total \,rates}$, and $N_{U}^{6b}=\sigma\times\epsilon_{6b}\times\epsilon_U\times L$ with $\epsilon_{6b} \equiv 0.58^3 \times 0.8^6$.  The event selection details are given in the main text. $N_{U}^{6b}$ displays the total number of events preserving unitarity with 6 $b$-jets, after the 3 Higgs boson decays.
\label{tab:lhc}}
\end{center}
\end{table*}

\begin{table*}[p]
\renewcommand{\arraystretch}{1.25} 
\begin{center}
  \begin{minipage}{.1\linewidth}
    \setlength{\tabcolsep}{0.2cm}
    \begin{tabular}{|c|c|}        
        \multicolumn{2}{c}{{}}\\
        \hline
        $b$ &  $c$  \\ \hline
        1.0 & 0.0 \\
        1.0 & 0.1 \\
        1.0 & 0.5 \\
        1.0 & 1.0 \\
        1.0 & 2.0 \\
        1.0 & 5.0 \\
        1.5 & 0.0 \\
        1.5 & 0.1 \\
        1.5 & 0.5 \\
        1.5 & 1.0 \\
        1.5 & 2.0 \\
        1.5 & 5.0 \\
        1.2 & 0.0 \\
        1.2 & 0.1 \\
        1.2 & 0.5 \\
        1.2 & 1.0 \\
        1.2 & 2.0 \\
        1.2 & 5.0 \\
        0.8 & 0.0 \\
        0.8 & 0.1 \\
        0.8 & 0.5 \\
        0.8 & 1.0 \\
        0.8 & 2.0 \\
        0.8 & 5.0 \\
        0.5 & 0.0 \\
        0.5 & 0.1 \\
        0.5 & 0.5 \\
        0.5 & 1.0 \\
        0.5 & 2.0 \\
        0.5 & 5.0 \\
        0.0 & 0.0 \\
        0.0 & 0.1 \\
        0.0 & 0.5 \\
        0.0 & 1.0 \\
        0.0 & 2.0 \\
        0.0 & 5.0 \\
        \hline
    \end{tabular}
    \end{minipage}
      \begin{minipage}{.4\linewidth}
                  \setlength{\tabcolsep}{0.35cm}
    \begin{tabular}{|S[table-format=1.3e-1]|S[table-format=1.3e-1]|c|}
        \hline
        \multicolumn{3}{|c|}{{$e^+e^-$, CLIC (3 TeV, 5 $\text{ab}^{-1}$)}}\\
        \hline
        {$\sigma$ (fb)} & {$\epsilon_{U}$} & {$N_{U}^{6b}$} \\ \hline
        2.56e-04 & 1.00e+00 &         0 \\
        1.68e-02 & 1.00e+00 &         4 \\
        4.08e-01 & 9.47e-01 &        99 \\
        1.68e+00 & 6.87e-01 &       294 \\
        6.59e+00 & 4.32e-01 &       728 \\
        4.17e+01 & 1.61e-01 &      1715 \\
        7.09e-01 & 8.73e-01 &       158 \\
        5.20e-01 & 8.86e-01 &       118 \\
        4.40e-02 & 1.00e+00 &        11 \\
        2.05e-01 & 9.91e-01 &        52 \\
        3.00e+00 & 6.03e-01 &       462 \\
        3.21e+01 & 2.42e-01 &      1984 \\
        1.14e-01 & 1.00e+00 &        29 \\
        4.44e-02 & 1.00e+00 &        11 \\
        9.53e-02 & 1.00e+00 &        24 \\
        9.13e-01 & 8.29e-01 &       194 \\
        5.04e+00 & 4.94e-01 &       637 \\
        3.74e+01 & 2.28e-01 &      2183 \\
        1.15e-01 & 1.00e+00 &        29 \\
        2.19e-01 & 9.86e-01 &        55 \\
        9.65e-01 & 8.25e-01 &       204 \\
        2.63e+00 & 6.49e-01 &       436 \\
        8.60e+00 & 4.71e-01 &      1036 \\
        4.61e+01 & 1.63e-01 &      1923 \\
        7.22e-01 & 8.58e-01 &       158 \\
        9.37e-01 & 7.91e-01 &       190 \\
        2.22e+00 & 6.63e-01 &       377 \\
        4.59e+00 & 4.91e-01 &       576 \\
        1.17e+01 & 3.40e-01 &      1019 \\
        5.32e+01 & 1.68e-01 &      2287 \\
        2.85e+00 & 5.73e-01 &       418 \\
        3.30e+00 & 5.77e-01 &       487 \\
        5.44e+00 & 4.82e-01 &       671 \\
        8.85e+00 & 3.40e-01 &       770 \\
        1.82e+01 & 2.92e-01 &      1355 \\
        6.61e+01 & 1.43e-01 &      2418 \\
        \hline
    \end{tabular}
  \end{minipage}
\caption{HEFT rates predicted with \mg~for triple Higgs production at future $e^+e^-$ collider  CLIC (3 TeV, $5~\text{ab}^{-1}$) for some selected values of $b$ and $c$.  The first row displays the SM rates. Here, $\sigma \equiv \sigma(e^+ e^-\rightarrow HHH \nu_e\bar{\nu}_e$),  $\epsilon_U \equiv {\rm rates\, preserving\, unitarity}/ {\rm total \,rates}$ and $N_{U}^{6b}=\sigma\times\epsilon_{6b}\times\epsilon_U\times L$ with $\epsilon_{6b} \equiv 0.58^3 \times 0.8^6$. The details and cuts are given in the text. $N_{U}^{6b}$ refers to the total number of events, preserving unitarity with 6 $b$-jets, after including the decay of the three Higgs bosons.
\label{tab:clic}}
\end{center}
\end{table*}

\begin{table*}[p]
\renewcommand{\arraystretch}{1.25} 
  \begin{minipage}{.1\linewidth}
            \setlength{\tabcolsep}{0.2cm}
    \begin{tabular}{|c|c|}        
        \multicolumn{2}{c}{{}}\\
        \hline
        $b$ & $ c$  \\ \hline
        1.0 & 0.0 \\
        1.0 & 0.1 \\
        1.0 & 0.5 \\
        1.0 & 1.0 \\
        1.0 & 2.0 \\
        1.0 & 5.0 \\
        1.5 & 0.0 \\
        1.5 & 0.1 \\
        1.5 & 0.5 \\
        1.5 & 1.0 \\
        1.5 & 2.0 \\
        1.5 & 5.0 \\
        1.2 & 0.0 \\
        1.2 & 0.1 \\
        1.2 & 0.5 \\
        1.2 & 1.0 \\
        1.2 & 2.0 \\
        1.2 & 5.0 \\
        0.8 & 0.0 \\
        0.8 & 0.1 \\
        0.8 & 0.5 \\
        0.8 & 1.0 \\
        0.8 & 2.0 \\
        0.8 & 5.0 \\
        0.5 & 0.0 \\
        0.5 & 0.1 \\
        0.5 & 0.5 \\
        0.5 & 1.0 \\
        0.5 & 2.0 \\
        0.5 & 5.0 \\
        0.0 & 0.0 \\
        0.0 & 0.1 \\
        0.0 & 0.5 \\
        0.0 & 1.0 \\
        0.0 & 2.0 \\
        0.0 & 5.0 \\
        \hline
    \end{tabular}
    \end{minipage}%
    \begin{minipage}{.4\linewidth}
                      \setlength{\tabcolsep}{0.35cm}
    \begin{tabular}{|S[table-format=1.3e-1,table-align-text-post=false]|S[table-format=1.3e-1]|c|}
        \hline
        \multicolumn{3}{|c|}{{$\mu^+\mu^-$ (3 TeV, 1 $\text{ab}^{-1}$)}}\\
        \hline
        {$\sigma$ (fb)} & {$\epsilon_{U}$} & {$N_{U}^{6b}$} \\ \hline
        2.56e-04 & 1.00e+00 &         0 \\
        1.68e-02 & 1.00e+00 &         1 \\
        4.08e-01 & 9.47e-01 &        20 \\
        1.68e+00 & 6.87e-01 &        59 \\
        6.59e+00 & 4.32e-01 &       146 \\
        4.17e+01 & 1.61e-01 &       343 \\
        7.09e-01 & 8.73e-01 &        32 \\
        5.20e-01 & 8.86e-01 &        24 \\
        4.40e-02 & 1.00e+00 &         2 \\
        2.05e-01 & 9.91e-01 &        10 \\
        3.00e+00 & 6.03e-01 &        92 \\
        3.21e+01 & 2.42e-01 &       397 \\
        1.14e-01 & 1.00e+00 &         6 \\
        4.44e-02 & 1.00e+00 &         2 \\
        9.53e-02 & 1.00e+00 &         5 \\
        9.13e-01 & 8.29e-01 &        39 \\
        5.04e+00 & 4.94e-01 &       127 \\
        3.74e+01 & 2.28e-01 &       437 \\
        1.15e-01 & 1.00e+00 &         6 \\
        2.19e-01 & 9.86e-01 &        11 \\
        9.65e-01 & 8.25e-01 &        41 \\
        2.63e+00 & 6.49e-01 &        87 \\
        8.60e+00 & 4.71e-01 &       207 \\
        4.61e+01 & 1.63e-01 &       385 \\
        7.22e-01 & 8.58e-01 &        32 \\
        9.37e-01 & 7.91e-01 &        38 \\
        2.22e+00 & 6.63e-01 &        75 \\
        4.59e+00 & 4.91e-01 &       115 \\
        1.17e+01 & 3.40e-01 &       204 \\
        5.32e+01 & 1.68e-01 &       457 \\
        2.85e+00 & 5.73e-01 &        84 \\
        3.30e+00 & 5.77e-01 &        97 \\
        5.44e+00 & 4.82e-01 &       134 \\
        8.85e+00 & 3.40e-01 &       154 \\
        1.82e+01 & 2.92e-01 &       271 \\
        6.61e+01 & 1.43e-01 &       484 \\
        \hline
    \end{tabular}
  \end{minipage}%
  \begin{minipage}{.42\linewidth}
                    \setlength{\tabcolsep}{0.35cm}
    \begin{tabular}{|S[table-format=1.3e-1]|S[table-format=1.3e-1]|c|}
        \hline
        \multicolumn{3}{|c|}{{$\mu^+\mu^-$ (10 TeV, 10 $\text{ab}^{-1}$)}}\\
        \hline
        {$\sigma$ (fb)} & {$\epsilon_{U}$} & {$N_{U}^{6b}$} \\ \hline
        3.20e-03 & 1.00e+00 &         2 \\
        2.93e+00 & 3.28e-01 &       491 \\
        7.33e+01 & 5.90e-02 &      2212 \\
        2.92e+02 & 3.40e-02 &      5083 \\
        1.17e+03 & 4.00e-03 &      2386 \\
        7.27e+03 & 6.00e-03 &     22308 \\
        1.29e+02 & 5.60e-02 &      3703 \\
        9.37e+01 & 4.70e-02 &      2254 \\
        7.93e+00 & 1.83e-01 &       742 \\
        3.34e+01 & 9.30e-02 &      1587 \\
        5.23e+02 & 2.60e-02 &      6961 \\
        5.44e+03 & 6.00e-03 &     16696 \\
        2.06e+01 & 1.22e-01 &      1284 \\
        7.77e+00 & 1.73e-01 &       687 \\
        1.61e+01 & 1.68e-01 &      1385 \\
        1.59e+02 & 4.40e-02 &      3573 \\
        8.88e+02 & 1.20e-02 &      5450 \\
        6.51e+03 & 2.00e-03 &      6654 \\
        2.04e+01 & 1.29e-01 &      1347 \\
        3.88e+01 & 8.50e-02 &      1686 \\
        1.71e+02 & 3.90e-02 &      3405 \\
        4.69e+02 & 1.60e-02 &      3837 \\
        1.50e+03 & 2.10e-02 &     16109 \\
        8.04e+03 & 2.00e-03 &      8222 \\
        1.28e+02 & 4.80e-02 &      3136 \\
        1.72e+02 & 3.50e-02 &      3071 \\
        3.95e+02 & 2.50e-02 &      5057 \\
        8.08e+02 & 1.90e-02 &      7855 \\
        2.08e+03 & 1.00e-02 &     10663 \\
        9.38e+03 & 6.00e-03 &     28776 \\
        5.06e+02 & 2.10e-02 &      5432 \\
        5.90e+02 & 1.80e-02 &      5429 \\
        9.46e+02 & 1.20e-02 &      5808 \\
        1.58e+03 & 7.00e-03 &      5649 \\
        3.25e+03 & 1.40e-02 &     23264 \\
        1.17e+04 & 4.00e-03 &     23906 \\
        \hline
    \end{tabular}
  \end{minipage}
\caption{HEFT rates predicted with \mg~for triple Higgs production at future $\mu^+\mu^-$ colliders, (3 TeV, $1~\text{ab}^{-1}$) (left subtable) and  (10 TeV, $10~\text{ab}^{-1}$) (right subtable) for some selected values of $b$ and $c$. The first row displays the SM rates. Here, $\sigma \equiv \sigma(\mu^+ \mu^-\rightarrow HHH \nu_\mu\bar{\nu}_\mu)$, $\epsilon_U \equiv {\rm rates\, preserving\, unitarity}/ {\rm total \,rates}$, and $N_{U}^{6b}=\sigma\times\epsilon_{6b}\times\epsilon_U \times L$ with $\epsilon_{6b} \equiv 0.58^3 \times 0.8^6$. The details and cuts are given in the text.  $N_{U}^{6b}$ displays the total number of events preserving unitarity with 6 $b$-jets, after the decay of the three Higgs bosons.
\label{tab:muo}}
\end{table*}

\section{Predictions for  future colliders}
\label{sec:future}
We now extend our discussion of the previous section to the prospects at future collider concepts for completeness. We collect the most relevant HEFT predictions of triple Higgs rates being produced mainly from WBF for $pp$, $e^+e^-$ and $\mu^+\mu^-$ machines (see also~\cite{Chiesa:2020awd,Han:2021lnp,Celada:2023oji} for related discussions). Concretely,   we explore 
\begin{figure}[!b]
\includegraphics[width=0.45\textwidth]{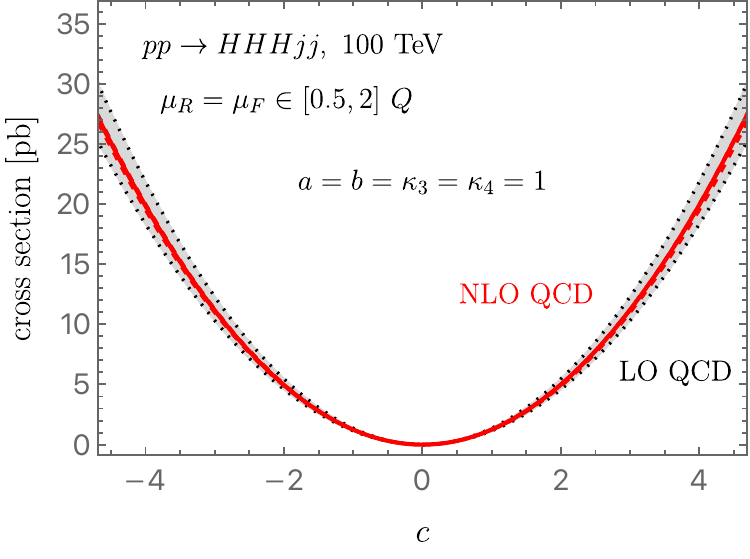}
\caption{\label{fig:xsec100} Cross section of $pp \to HHHjj$ as function of $c$ (cf.~Fig.~\ref{fig:xsec}) for 100~TeV collisions.}
\end{figure}
\begin{enumerate}[start=0]
    \item HL-LHC ($pp$, 14~TeV, $3~\text{ab}^{-1}$),
    \item FCC-hh ($pp$, 100~TeV,  $30~\text{ab}^{-1}$),
    \item CLIC ($e^+e^-$,  3~TeV,  $5~\text{ab}^{-1}$),
    \item  $\mu^+\mu^-$,  3 TeV,  $1~\text{ab}^{-1}$,
    \item  $\mu^+\mu^-$,  10 TeV,  $10~\text{ab}^{-1}$.
\end{enumerate}
To compare the sensitivity to the $c$ parameter in the two chosen $pp$ colliders, we show in Fig.~\ref{fig:xsec100} the cross sections for $pp \to HHHjj$ for 100 TeV collisions similar to Fig.~\ref{fig:xsec}. Overall, we find a cross section increase of about $350\times \sigma(\text{14~TeV})\simeq \sigma(\text{100~TeV})\simeq 75~\text{ab}$ for SM coupling choices. The qualitative features discussed in Sec.~\ref{sec:lhc} for the 14 TeV case carry over to the FCC-hh case, i.e. the systematic limitations due to QCD are comparably irrelevant.

\newenvironment{cedescription}{%
	\renewcommand\descriptionlabel[1]{\hspace{\labelsep}{{\textit{##1}}}}
	\begin{description}[leftmargin=0.25cm, style=sameline]%
	}{%
	\end{description}%
}
\newcommand{\BR}{\text{BR}}

A summary of the HEFT rates in relation to our discussion unitarity in the previous section is given in Tabs.~\ref{tab:lhc},~\ref{tab:clic}, and~\ref{tab:muo} for the FCC-hh, CLIC and the staged muon collider.
\begin{cedescription}
\item[0.) HL-LHC and 1.) FCC-hh]: For the $pp$ collider cases, we consider $pp \to HHHjj$ using the modified version of {\tt{vbfnlo3}} already used in Sec.~\ref{sec:lhc}. The cuts applied on the WBF pair $jj$ jets are as in Eq.~\eqref{eq:cuts}  The reduction factor from unitarity constraints $\epsilon_U$ is defined as in Eq.~\ref{epsilonU}. The final state with 6$b$-jets  and the two WBF jets, $6b+jj$ is approximated from assuming the three Higgs bosons decaying into $b \bar b$ pair centrally. The branching ratio ($\BR$) for each such Higgs boson decay is 0.58~\cite{LHCHiggsCrossSectionWorkingGroup:2011wcg}. $N_{U}^{6b}$ represents the total number of events preserving unitarity with 6$b$-jets, after Higgs decays. We have assumed a tagging $b$ jet efficiency of 80\%~\cite{ATLAS:2012ima}. Hence, the predicted final events, before additional analysis efficiencies, is by 
\begin{equation}
\label{eq:nu}
N_{U}^{6b}=\sigma\times{}\epsilon_{6b}\times{}\epsilon_U\times{}L~\text{with}~\epsilon_{6b} \equiv 0.58^3 \times 0.8^6\,.
\end{equation}
\item[2.) CLIC]: To produce the desired triple Higgs final state, we have to rely on WBF production at the highest energies, hence we consider CLIC. For this specific high-energy $e^+e^-$ case, the full process studied is $e^+ e^-$$\rightarrow $$HHH \nu_e\bar{\nu}_e$ using~\mg. We cut on the missing transverse energy from the final $\nu \bar \nu$ pair by imposing $\slashed{E}_T> 20~\text{GeV}$. The reduction factor from unitarity constraints $\epsilon_U$ is again defined by Eq.~\eqref{epsilonU}. The final state with 6 $b$-jets in addition to missing energy,  $6b+\slashed{E}_T$, again assuming the three Higgs bosons decaying into the dominant $b \bar b$ channel with $\BR=0.58$. $N_{U}^{6b}$ represents the total number of events preserving unitarity with 6$b$-jets with efficiency of 0.8 for comparability. The predicted final events are given by $N_{U}^{6b}$ as above.
\item[3.) and 4.) staged muon collider]: For the $\mu^+ \mu^-$ machine, similar to CLIC, we consider $\mu^+ \mu^-\rightarrow HHH \nu_\mu\bar{\nu}_\mu$ using~\mg5, again for $\slashed{E}_T> 20~ \text{GeV}$ (a more comprehensive analysis of this process has been presented recently in~\cite{Celada:2023oji}). Reduction factors, branching ratios, $b$ tagging efficiencies are left unchanged compared to the FCC and CLIC, therefore $N_U^{6b}$ is again given by Eq.~\eqref{eq:nu}
\end{cedescription}

The general features of the comparison of Tabs.~\ref{tab:lhc},~\ref{tab:clic}, and~\ref{tab:muo} is that all predicted triple Higgs cross sections for the BSM scenarios with $(b,c) \neq (1,0)$ depart considerably from the SM rates as expected from the WBF phenomenology detailed in Sec.~\ref{sec:lhc}. The search for BSM effects from non-standard contact interactions $VVHH$ and $VVHHH$ (cf.~Sec.~\ref{sec:intro}), albeit highly challenging at the LHC, presents an opportunity for future colliders. 
That said, the restrictions from unitarity requirements are considerably larger at the higher energy colliders. As these ultimately underpin the current methodolgy of limit setting, these are relevant constraints that require consideration when moving to ${\cal{O}}(10~\text{TeV})$ parton colliders. The most stringent restrictions are found for FCC-hh (100TeV)  and  $\mu^+\mu^-$ (10~TeV) collisions, where reduction factors of $\epsilon_U \sim 10^{-3}-10^{-2}$ signal potential self-consistency issues (we stress again that this will depend on whether the relevant scale of unitarity is resolved~\emph{experimentally}). Comparably less affected by unitarity considerations are lepton colliders with 3~TeV.\footnote{We note when comparing the results in Table \ref{tab:muo} (for 3 TeV muon collider) and Table \ref{tab:clic} (for 3 TeV electron collider), we get similar results, as expected, except for the trivial factor 5 due to the different luminosities.}

From the last column in these Tabs.~\ref{tab:lhc}-\ref{tab:muo}, we conclude that all these future colliders will offer a significant exploration potential, going well beyond the current  LHC forecasts. Of course, this extends to studies beyond the rare SM multi-Higgs weak boson fusion processes considered here, but they contribute to the motivation of further developing the physics profile of any future machine in more experimentally realistic case studies. Finally, electroweak corrections at large momentum transfers can significantly alter the leading-order electroweak phenomenology we discuss here. For the Standard Model, these have been discussed in, e.g.,~\cite{Ciccolini:2007jr,Ciccolini:2007ec,Pagani:2021vyk, Ma:2024ayr} for the LHC and beyond. In particular, the high-energy lepton environment probed at CLIC or a future muon collider will undoubtedly affect the leading order exclusion limits quoted in this work. Results are not currently available in full, and this will become a pressing question when the future collider roadmap has been clarified.

\section{Conclusions}
\label{sec:conc}
Multi-Higgs production is becoming increasingly interesting to the LHC experimental collaborations with double Higgs processes getting under increasing statistical control~\cite{ATLAS:2023qzf,CMS:2022hgz}. The next natural step in the exploration of the electroweak scale at high-energy colliders is, therefore, triple Higgs production. 
Relevant processes are gaining experimental and theoretical attention, particularly in relation to the preparation of future collider concepts. 

In this work, we have considered WBF triple Higgs production within the context of HEFT,  with a view towards weak gauge-Higgs contact interactions that are heavily suppressed in the SM and its perturbative EFT extension. From the SM's perspective, such an analysis, therefore, appears futile. Considering a more general parameterisation of the electroweak vacuum consistent with the current LHC physics programme, there is, however, no reason to dismiss these: We have shown that WBF triple Higgs production provides a QCD-stable window into properties of the electroweak sector, for example when viewed through the lens of scalar sector geometry~\cite{Alonso:2016btr,Alonso:2021rac}. Although there are no unitarity-based no-go theorems in high-energy electroweak physics after the Higgs boson discovery, a more detailed exploration of the very dynamics of electroweak symmetry breaking is at a relatively early stage at the LHC. This leaves ample opportunity to observe non-standard interactions in the interactions of multiple Higgs bosons which are known to probe different aspects of electroweak physics compared to single Higgs observations. 

These contact interactions are particularly highlighted by non-SM contact terms $\sim VVH^3$, which are naturally parametrised using Higgs Effective Field Theory. Phenomenologically, these terms can be relevant as they open an avenue for significant cross sections balanced against BSM unitarity requirements. We have analysed these processes' sensitivity against these self-consistency requirements to highlight non-negligible (and QCD-stable) BSM parameter regions that can be explored at TeV-scale colliders. Here, we have analysed both hadron and lepton colliders, particularly,  LHC, FCC, CLIC and muon colliders. Hence, the search for double and triple Higgs production at present and future colliders will clarify the currently unknown dynamics responsible for electroweak symmetry breaking.

\bigskip 
\noindent {\bf{Acknowledgements}} ---
\noindent A. is funded by the Leverhulme Trust under Research Project Grant RPG-2021-031. 
The work of D.D. is also supported by the Spanish Ministry of Science and Innovation via an FPU grant No FPU22/03485.
D.D. and M.J.H. acknowledge financial support from the grant IFT Centro de Excelencia Severo Ochoa CEX2020-001007-S funded by MCIN/AEI/10.13039/ 501100011033, from the Spanish ``Agencia Estatal de Investigaci\'on'' (AEI) and the EU ``Fondo Europeo de Desarrollo Regional'' (FEDER) through the project PID2019-108892RB-I00 funded by MCIN/AEI/ 10.13039/501100011033, and from the European Union’s Horizon 2020 research and innovation programme under the Marie Sklodowska-Curie Grant agreement No. 860881-HIDDeN. D.D. and M.J.H. also acknowledge partial financial support by the Spanish Research Agency (Agencia Estatal de Investigaci\'on) through the Grant PID2022-137127NB-I00 funded by MCIN/AEI/10.13039/501100011033/FEDER, UE. 
C.E. is supported by the UK Science and Technology Facilities Council (STFC) under grant ST/X000605/1, the Leverhulme Trust under Research Project Grant RPG-2021-031 and Research Fellowship RF-2024-300/9, and the Institute for Particle Physics Phenomenology Associateship Scheme.
The work of R.A.M. is supported by CONICET and ANPCyT under project PICT-2021-00374.

\bibliographystyle{apsrev4-1}
\bibliography{paper-PRDv4}

\end{document}